\documentclass{elsarticle}

\usepackage{amsmath}
\usepackage{epsf}
\usepackage{graphicx}
\usepackage{dcolumn}
\usepackage{bm}

\newcommand{\req}[1]{Eq.~(\ref{#1})}

\newcommand{\rref}[1]{(\ref{#1})}

\newcommand{\beq}{\begin{equation}}
\newcommand{\eeq}{\end{equation}}
\newcommand{\be}{\begin{equation}}
\newcommand{\ee}{\end{equation}}
\newcommand{\beqa}{\begin{eqnarray}}
\newcommand{\eeqa}{\end{eqnarray}}
\newcommand{\bea}{\begin{align}}
\newcommand{\eea}{\end{align}}
\usepackage{amssymb}
\usepackage{array}
\usepackage{amsmath}
\usepackage{graphicx}\usepackage{graphics}
\usepackage{dcolumn}
\usepackage{bm}\usepackage{varioref}

\usepackage{amsmath}
\usepackage{epsf}
\usepackage{graphicx}
\usepackage{subfigure}
\usepackage{dcolumn}
\usepackage{bm}

\usepackage{amssymb}
\usepackage{array}
\usepackage{varioref}
\usepackage{wrapfig}
\usepackage{etoolbox}
\usepackage{color}

\usepackage{esint}

\begin{document}

\title{
 Saturation of strong electron-electron umklapp scattering at high temperature}

\author[h]{Igor L. Aleiner\footnote{Corresponding author. Email:
aleiner@phys.columbia.edu}}
\address[h]{Physics Department, Columbia University, New York, NY 10027, USA}
\author[c]{Oded Agam}
\address[c]{The Racah Institute of Physics, The Hebrew University of Jerusalem, 91904, Israel}
%\date{\today}
\begin{abstract}
  We consider clean metals, at finite temperature, in which the
  inelastic rate, $\hbar/ \tau_{ee}$, can become of the order of, or
  larger, than the band splitting energy. We show that in such
  systems, contrary to the common knowledge, the umklapp scattering
  rate becomes independent of both $\tau_{ee}$ and the
  temperature $T,$ in three dimensional systems. We discuss the relation of this phenomenon to the
  saturation of resistivity at high temperature.

\end{abstract}
\maketitle

\section{Introduction}
 
Electron-electron umklapp scattering is
the process by which momentum relaxation of electrons
occurs in clean systems, by transferring it to the lattice.
It is believed that the rate of umklapp scattering is always proportional to the inelastic rate $1/\tau_{ee}$ \cite{Landau1937}. In the nearly free electron limit,
umklapp scattering can be viewed as a two stage process: first, an electron
excites an electron-hole pair into a virtual state, and then, one of these particles is scattered by the
periodic potential of the lattice. If there are no singularities in the corresponding matrix element,  the rate of this process can
be deduced from the argument of phase-space volume, leading
to the well known $1/\tau_{ee}$ dependence. In particular, for Fermi liquid systems,$1/\tau_{ee}\propto T^2$.

In this paper we show that, contrary to this
general belief, in certain cases, the umklapp scattering rate saturates
to a finite value which is independent of $\tau_{ee}$.
It occurs when the Fermi level lies near the boundary of the Brillouin zone (BZ). In this case,  the aforementioned intermediate state can be
at resonance. The singularity of the corresponding matrix element is cutoff by the inelastic rate itself, leading to an  umklapp scattering rate  which is independent of $\tau_{ee}$, and hence independent of the temperature, see \req{main}.

The motivation for this study  comes from the long standing problem of the so-called high temperature saturation of resistance in metals (which appear in many metals and mostly in $A15$ compounds)\cite{Woodard64,Marchenko73,Fisk76,Testardi76,
Wiesmann77,Gurvitch78,Gurvitch80,Oota80,Oota81}.  In early works on the subject \cite{review}, the effect was ascribed to "localization" in a random field created by the phonons, thereby claiming the saturated resistivity value to be the "maximal metallic resistivity", $\rho_{max}\sim \frac{2\pi \hbar}{e^2} \cdot 1\mbox{\AA}\sim 300 \mu \Omega \cdot \mbox{cm}$, without any microscopic grounds. This value is known as the Ioffe-Regel limit \cite{IoffeRegel60}.  Here we show that resistivity saturation, in cases where Fermi surface touches the boundary the BZ with several bands (and $A15$ compounds are indeed such materials \cite{A15Bands}), is due to the saturation of the umklapp scattering, and is not sensitive to the phonon physics.

\begin{figure}[t]
\includegraphics[width=0.98\columnwidth]{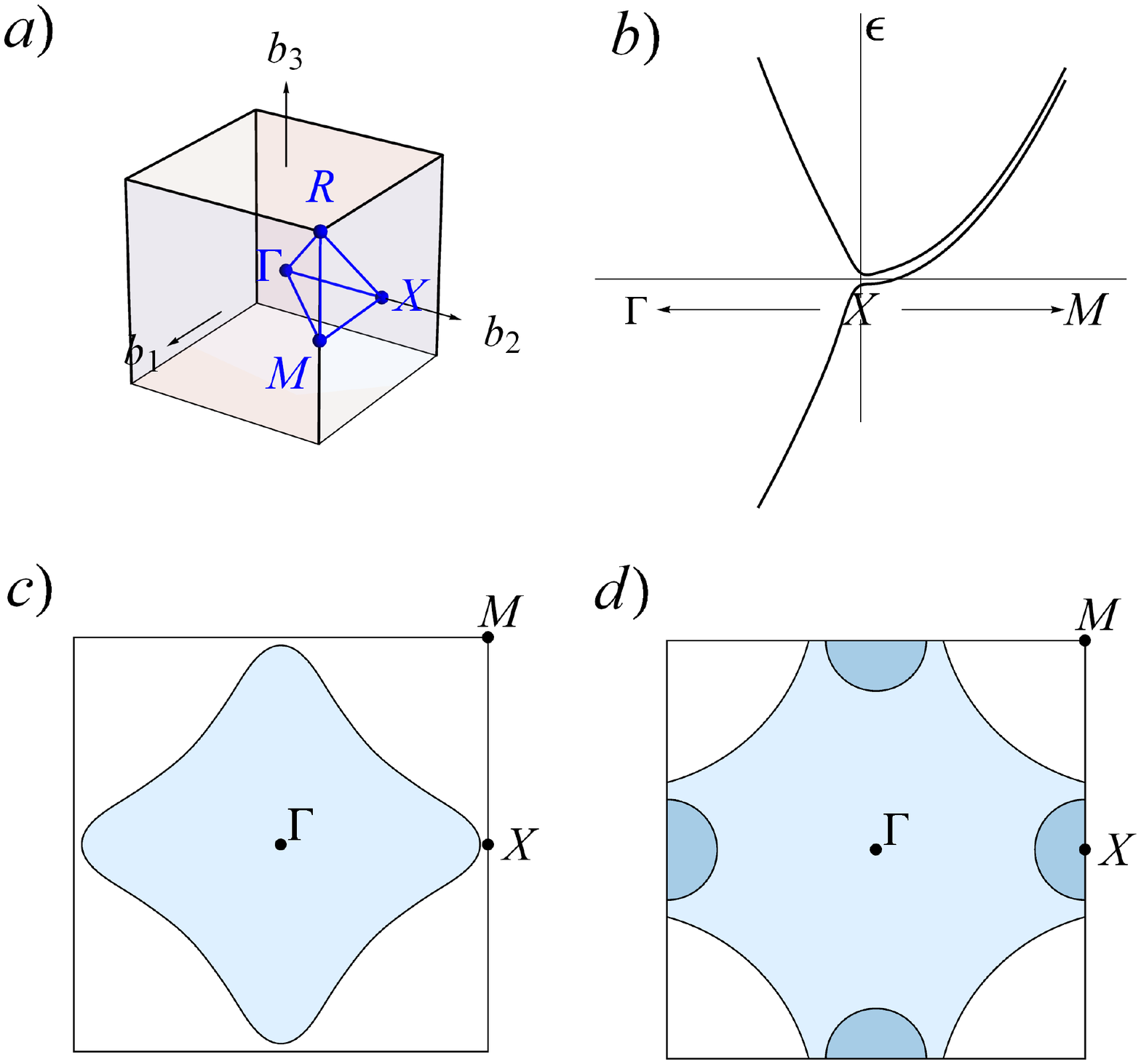}
\caption{a) The Brillouin zone of a simple cubic lattice and its symmetry points. b) An Illustration of the energy bands near X-point of a simple cubic lattice. c) and d) Fermi surfaces forming a "hot spot"  and  a "hot line", respectively, for resonant umklapp scattering near the edge of the Brillouin zone.}
\label{fig1}
\end{figure}

We consider a three dimensional system described by the Hamiltonian
\be
H= H_{h}+ \int d^3r U({\bf r})\psi^{\dagger}({\bf r}) \psi({\bf r}). 
\ee
Here $H_{h}$ stands for the Hamiltonian of an interacting electronic system 
(and characterized by a typical inelastic rate $1/\tau_{ee} <T $, we will use $\hbar=1$ from now on and restore $\hbar$ only in final formulas.)
which conserves momentum (does not contain any umklapp process), while
\be 
U({\bf r})=\sum_{{\bf b}} U_{{\bf b}} e^{ i {\bf b} \cdot {\bf r}}~~; ~~~   U_{\bf b}= U^*_{-{\bf b}} \label{eq:U}
\ee
 is a weak periodic potential (${\bf b}$ being a vector of the reciprocal lattice) which at zero temperature generates  energy splitting as shown in Fig.~1b. Finally, $\psi({\bf r})$ and  $\psi^\dagger({\bf r})$ are Fermionic fields, and the sum over spin indices is implied.
All the results in this paper are obtained assuming that $|U| <T $ but for the arbitrary relation between $U$ and $1/\tau_{ee}$.

 A qualitative understanding for the resistivity saturation in this
 model follows from consideration of Bloch wave function on the level
 of a single quasi-particle. Bloch wave functions result from the
 constructive interference of multiple scattering of an electron by
 the periodic potential. The typical time scale for this process is
 the inverse energy gap opened by the potential, $\tau_U\simeq 1/|U|$.  However,
 the scattering of the electron by other electrons introduces a cutoff
 on this process at time of order $\tau_{ee}$.  Thus when $\tau_{ee}
 \gg \tau_U$, the electron oscillates in the periodic potential, the band structure is formed, and the periodic potential itself does not control the resistance.
The latter is determined by deviation of the potential from the periodicity  due to defects or phonon field (which can be considered as static at temperatures larger than the Debye 
temperature). In the clean metals deviation of the potential from the periodicity is always smaller than the periodic potential itself due to Lindemann criteria\cite{Lindemann1910}.

In the other limit, $\tau_{ee} \ll \tau_U$, scattering by other electrons prevents the construction of Bloch wave
 from the interference of multiple scattering, altogether. The
 scattering of the electron by the periodic potential of the lattice
 becomes independent of $\tau_{ee}$, and plays the role similar to scattering by the static
 disorder potential.   In this case,  phonons and disorder play only a minor role when saturation is reached.

\section{Hydrodynamic equations}
Due to the strong inelastic electron-electron interaction which conserves momentum, the appropriate  description of the system is hydrodynamics, 
and we start by summarizing this description 
(the hydrodynamic description of transport in metals at low temperatures was developed quite long ago,
see Ref.~\cite{Gurzhi} for the review. Our results at high temperature differ in microscopic expression for the force density ${\bf f}_U$, cf.~Eqs.~(\ref{eq:ff})). 

We introduce the electron density $n$,  the velocity field ${\bf v}$, and the thermodynamic energy density $\epsilon(n,s,{\bf p})$, which is a function of the density, the entropy density $s$, and the momentum density ${\bf p}$.
The latter satisfies the thermodynamic relation:
\be
d\epsilon=Tds+\mu dn+ v_\alpha d p_\alpha,
\ee
where $T$ is the temperature, $\mu$ is the chemical potential, and, finally,  the pressure is given by
\be
P=-\epsilon+ \mu n + Ts+ {v}_\alpha  { p}_\alpha.
\ee

\begin{subequations}
The density satisfies the continuity equation:
\be
\partial_t n + \partial_{\alpha} (n  v_\alpha) =0. \label{eq:n}
\ee
The momentum balance equation is
\be
\partial_t p_\alpha+ \partial_{\beta} (v_\beta p_\alpha)+\partial_\alpha P = en E_\alpha + f^{U}_\alpha - \partial_\beta \mathbb T^{(ne)}_{\beta \alpha}, \label{eq:p}
\ee
where ${\bf E}$ is the applied electric field, ${\bf f}^{U}$ is the force density emerging from the periodic potential (\ref{eq:U}) which will be calculated below, and $\mathbb T^{(ne)}$ is the contribution to the stress tensor due to the non-equilibrium distribution function (i.e. the part of the distribution function which is not a zero mode of the collision integral). More details about the hydrodynamic description of non-Galilean invariant systems and crystals with weak umklapp scattering can be found in the Appendix.

Following the standard route we replace energy conservation  by the entropy  balance equation:
\be
\partial_t s+ \partial_\alpha q_\alpha\! =\!-\frac{1}{T} \left( {\bf v}\cdot {\bf f}^U + \delta {\bf j}^\epsilon \cdot \frac{ {\bf \nabla} T}{T} +\mathbb T_{ik}^{(ne)} \partial_i v_k \right), \label{eq:s}
\ee
where 
\be 
q_\alpha= v_\alpha s+ \frac{\delta j_\alpha^\epsilon}{T} \label{eq:q}
\ee
\end{subequations}
is the entropy flux, while $\delta {\bf j}^\epsilon$ is the non-equilibrium correction to the energy current.

The entropy balance equation (\ref{eq:s}) allows one to identify the matrix of the kinetic coefficients which satisfies the Onsager relations. In the limit ${\bf v} \to 0$ there can be no proportionality between tensors and vectors, therefore:
\begin{subequations}
\be
\mathbb T_{\alpha\beta}^{(ne)} =- \nu_{\alpha \beta; \alpha'\beta'} \partial_{\alpha'} v_{\beta'} , 
\ee
where $\nu_{\alpha\beta;\alpha'\beta'}=\nu_{\alpha'\beta';\alpha\beta}$ is the viscosity tensor, while
\be
\left( \begin{array}{c} \delta j_\alpha^{\epsilon} \\ f_\alpha^U \end{array} \right) = -\left( \begin{array}{cc} \kappa_{\alpha\beta}T & \gamma_{\alpha\beta} \\ \gamma_{\beta\alpha} & \eta_{\alpha\beta} \end{array} \right)\left( \begin{array}{c} \frac{ \partial_\beta T}{T} \\  v_\beta \end{array} \right). \label{OR2}
\ee
\end{subequations}
Here $\hat{\kappa}$ is thermal conductivity tensor, $\hat{\eta}$ is the friction tensor, while $\hat{\gamma}$ is a tensor which should be neglected, as we argue below. Notice that both the viscosity tensor and the matrix of tensors on the right-hand-side (r.h.s.) of the last equation are positive definite in order to ensure entropy growth.

To see that $\hat{\gamma}$ can be neglected, we notice that being a coefficient that describes non-equilibrium quantities it must be proportional to the inelastic relaxation time, $\tau_{ee}$, as well as to $U^2$, i.e. $\hat{\gamma}\sim \tau_{ee}U^2$. On the other hand, the correction to the entropy density due to the periodic potential is $\delta s\sim s U^2/T^2$ (and does not depend on the relaxation time $\tau_{ee}$). Since we consider the limit   where $1/\tau_{ee} \leq T$, taking into account $\hat{\gamma}$ would be overstepping  of accuracy, 
and has to be neglected.

Assuming stationary state,  it follows from  Eqs. (\ref{eq:p}) (neglecting nonlinear terms), (\ref{eq:q}), (\ref{OR2}), and the thermodynamic relation $dP= \mu dn+ s dT$ that:
\be
\left( \begin{array}{c} {\bf E}- \frac{{\bf \nabla} \mu}{e} \\ {\bf q} \end{array} \right) = \left( \begin{array}{cc} \hat{\rho} & \frac{\Pi}{T} \\\frac{\Pi}{T} & -\frac{\hat{\kappa}}{T} \end{array} \right)\left( \begin{array}{c} {\bf j} \\  {\bf \nabla}T \end{array} \right),
\ee 
where ${\bf j}= ne{\bf v}$ is the electric current density, 
\be
\hat{\rho}=   \frac{ \hat{\eta}}{e^2 n^2} \label{eq:rho}
\ee
is the resistivity tensor, and
\be
\Pi=\frac{Ts}{en}
\ee
is the Peltier coefficient, which depends on the ``entropy per carrier'' 
\cite{Book}.
This formula shows that in the hydrodynamic limit the Peltier coefficient is expressed only in terms of  thermodynamic quantities. Therefore,
 the relation to the specific heat $C_v$ (from which the constant phonon contribution is subtracted), $d (\Pi/T)/d\ln T = C_v/en$, allows one to compare the value of $\Pi$ obtained from transport measurements 
to the result of thermodynamic measurements.

\section{Microscopic  calculation of the friction force density ${\bf f}^U$.} The starting point for the calculation of the force is its definition:
\be 
f_\alpha^U=- \langle\langle \psi^\dagger({\bf r}) \psi({\bf r}) \partial_\alpha U({\bf r}) \rangle_{{\bf v}}\rangle_{{\bf r}}, \label{def:f}
\ee
where averaging is over the thermodynamic state with velocity ${\bf v}$, and over space. Introducing the Keldysh Green function, $G^K_{{\bf v}}$, associated with velocity ${\bf v}$,  Eq.~(\ref{def:f}) may be rewritten in the form \cite{KamenevBook}:
\be 
f_\alpha^U=i \left\langle  \int \frac{ d\epsilon}{2\pi} G_{\bf v}^K(\epsilon; {\bf r},{\bf r}) \partial_\alpha U({\bf r})\right\rangle_{{\bf r}}, \label{eq:f2}
\ee 
where double spin degeneracy is taken into account.

It will be convenient to express the Keldysh Green function in terms of the retarded and advanced Green functions calculated at ${\bf v}=0$:
\be
\hat{G}^{R,A}= \left[ \epsilon-\hat{\xi}^{R,A}(\epsilon) -\hat{U}^{R,A}(\epsilon) \right]^{-1} 
\ee
where $[ \cdots]^{-1}$ should be understood as matrix inversion, and  
\be
\hat{\xi}^{R,A}(\epsilon) = \left[\xi({\bf p})+ \Sigma^{R,A}(\epsilon, {\bf p})\right]\delta_{{\bf p}, {\bf p}'}
\ee
characterize the single particle spectrum of $H_h.$ Here $\xi({\bf p})$ is the bare energy spectrum of the non-interacting counterpart of $H_h$, and $\Sigma^{R,A}(\epsilon, {\bf p})$ are the retarded and advanced  self energies.  Finally, $\hat{U}^{R,A}$ is a matrix whose elements are given by a sum over the reciprocal lattice vectors:  $\sum_{{\bf b}}U^{R,A}_{{\bf p}, {\bf p}'} \delta_{{\bf p}-{\bf p}', {\bf b}}$, where $U^{R,A}_{{\bf p}, {\bf p}'}$ are the periodic potential vertices dressed to all orders in the interaction, see Fig.~2.

In what follows it will be convenient to represent the Keldysh Green function in the form
\be 
\hat{G}^K= \hat{G}^R\hat{n}(\epsilon)- \hat{n}(\epsilon)\hat{G}^A, \label{eq:GK}
\ee 
where $\hat{n}(\epsilon)$ (not to be confused with the density, $n$) is a matrix in ${\bf p}$ space which represents the generalized distribution function. In equilibrium, it is given by $\hat{n}_{eq}(\epsilon)=n_0(\epsilon)\delta_{{\bf p},{\bf p}'}$ where $n_0(\epsilon)= 1- 2f_F(\epsilon)=\tanh[(\epsilon-\mu)/(2T)]$ is related to the Fermi distribution function, $f_F(\epsilon)$. Out of equilibrium, when ${\bf v} \neq 0$, the generalized distribution function satisfies the stationary equation \cite{GDF}:
\begin{subequations}
 \label{eq:nhat}
\begin{eqnarray}
&&i[\hat{H};\hat{n}]=\mbox{St}_{ee}\{\hat{n} \}+ \mbox{St}_U\{\hat{n} \}
\end{eqnarray}
where
\begin{eqnarray}
&&\mbox{St}_{ee}\{\hat{n} \}= iZ \left[\frac{1}{2}\{ \hat{\Sigma}_{{\bf v}} ^A- \hat{\Sigma}_{{\bf v}}^R;\hat{n}\}+ \hat{\Sigma}^K_{{\bf v}}\right],\\
&&\mbox{St}_U\{\hat{n} \}= iZ \left[\frac{1}{2} \{\hat{U}_{{\bf v}}^A-\hat{U}_{{\bf v}}^R;\hat{n}\}+\hat{U}^K_{{\bf v}}\right].\label{stu}
\end{eqnarray}
Here square [ ; ] and curly brackets \textbraceleft\ ; \textbraceright\ denote commutator and anti-commutator, respectively,
$Z= (1- \mbox{Re} \frac{d \Sigma^R}{d\epsilon})^{-1}$ is the quasi-particle  weight, $\hat{\Sigma}^K$ and $\hat{U}^K$ are the Keldysh self energy and potential respectively, the subscript ${\bf v}$ implies evaluation of the quantity  at finite velocity, and
\begin{eqnarray}
&&\hat{H}=\hat{\zeta}_{{\bf v}}+ \hat{{\mathbb U}}_{{\bf v}},
\\
&&\hat{\zeta}_{{\bf v}}= \frac{Z}{2} \left(\hat{\xi}^R_{{\bf v}}+  \hat{\xi}^A_{{\bf v}}\right),~~~~\hat{{\mathbb U}}_{{\bf v}}=\frac{Z}{2}\left( \hat{U}^R_{{\bf v}}+ \hat{U}^A_{{\bf v}}\right).  \nonumber
\end{eqnarray}
\end{subequations}
At equilibrium, fluctuation dissipation theorem implies $\Sigma^K= (\Sigma^R-\Sigma^A)n_0(\epsilon)$,  and  $U^K= (U^R-U^A)n_0(\epsilon)$.
 
\begin{figure}[h]
\includegraphics[width=0.7\columnwidth]{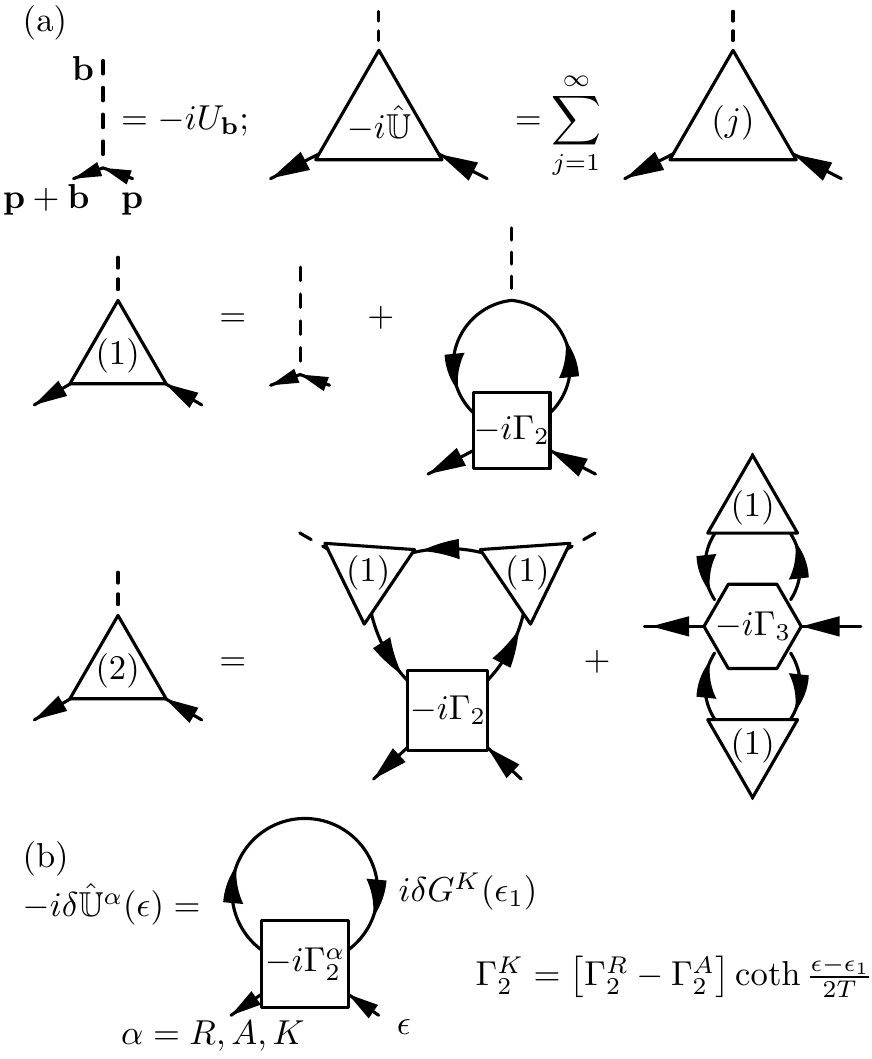}
\caption{(a) Diagrammatic expansion for equilibrium potentials  $\hat{U}^R$ and $\hat{U}^A$, where $\Gamma_j$ is the exact $j$-th particle vertex function for the system described by $H_h$. Notice that only graphs with finite momentum transfer are included, as the momentum conserving ones belong to the self energy.  (b) The perturbation of the effective potential due to the change of the distribution function. For small energy transfer, the vertex can be recast in the form 
$Z^2\mbox{Im}\Gamma_2^A =  \gamma\left( \frac{ \epsilon-\epsilon_1}{v_f q}\right)/\nu$, see text following Eq.~(\ref{dUs}).}
\end{figure}

Using Eqs. (\ref{eq:f2}), (\ref{eq:GK}), and the diagrammatic expansion of Fig.~2, we express the force in the form
\be
{\bf f}^U\!\!= \mbox{Tr}_{{\bf p}}\! \int \frac{ d\epsilon}{2\pi} \left( [\hat{U}^R;\hat{\bf p}]\hat{G}^R \hat{n}(\epsilon)-  \hat{n}(\epsilon)  \hat{G}^A[\hat{U}^A;\hat{\bf p}] \right) \label{eq:f3}
\ee   
Notice that our choice of using fully dressed potentials instead of the original ones means that the effect of finite velocity has to be taken only in 
$\hat{n}$ and not in $G^{R,A}$.

At zero velocity, ${\bf v}=0$, the friction force vanishes, ${\bf f}^U=0$,  and the Green function can be found using thermal equilibrium diagrammatic technique with proper analytic continuation. Finite velocity is imposed by the constraint:   
\be
n {\bf v}= \int \frac{ d\epsilon}{2\pi i} \left[ \frac{\partial \hat{\xi}^R}{\partial {\bf p}}
\hat{G}^R \hat{n} - \hat{n} \hat{G}^A \frac{\partial \hat{\xi}^A}{\partial {\bf p}}\right]. \label{eq:nv}
\ee 

 The generalized distribution function can be represented in the form $\hat{n}(\epsilon)=\hat{n}_{eq}(\epsilon) + \delta \hat{n}(\epsilon)$,
with the non-equilibrium contribution, $\delta \hat{n}=\delta \hat{n}_1+\delta \hat{n}_2+\delta \hat{n}_3+ \cdots$, that will be calculated by iterations of Eqs.~(\ref{eq:nhat}), starting with 
\be
\delta \hat{n}_1(\epsilon) =-{\bf p} \cdot {\bf v}\partial_\epsilon \hat{n}_{eq}(\epsilon).
\label{first}
\ee
One can verify, using integration by parts, that $\delta \hat{n}_1$  satisfies the constraint of Eq.~(\ref{eq:nv}).
Then, substituting $\hat{n}(\epsilon)=\hat{n}_{eq}(\epsilon) + \delta \hat{n}(\epsilon)$ in  Eq.~(\ref{eq:nhat}), expanding the self energies and the potentials in terms of  $\delta \hat{n}_1(\epsilon)$, and taking into account  momentum conservation, \be
\mbox{St}_{ee}\{ \hat{n}_{eq}+ \delta \hat{n}_1 \}=0, \label{eq:SK1}
\ee
we obtain that the non-equilibrium contribution to the generalized distribution function can be obtained from iterative  solution of the equation:
\begin{eqnarray}
&&i[H, \delta \hat{n}_{j+1}]-\mbox{St}_{ee}\{\hat{n}_{eq}+\delta \hat{n}_{j+1}\} \label{iterations}\\
&&~~~~~~~~~~=-i[H, \delta \hat{n}_1]\delta_{j,1}+\mbox{St}_U\{ \hat{n}_{eq}+\delta \hat{n}_j \}.  \nonumber
\end{eqnarray}
 Notice that the effect of the zero mode  in the collision integral $\mbox{St}_{ee}\{ \hat{n}_{eq}+ \delta \hat{n} \}$ is removed by  the constraint (\ref{eq:nv}), similarly to the Chapman-Enscog procedure \cite{Enscog}, and may result only in a perturbative renormalization of the velocity ${\bf v}$. 

{\it The two band model:} To be concrete, we consider a two band model with spectrum as shown in Fig.~1. This model describes the behavior near the X-point at the edge of the BZ of a simple cubic lattice. Nevertheless, as we explain later, it also captures the behavior of other Fermi surfaces on the boundary of the BZ. 

For the two band model, it is convenient to double the matrices in the band space:
\be
A \to \left( \begin{array}{cc} A_{{\bf p}+\frac{{\bf b}}{2},{\bf p}+\frac{{\bf b}}{2}} &  A_{{\bf p}+\frac{{\bf b}}{2},{\bf p}-\frac{{\bf b}}{2}} \\  A_{{\bf p}-\frac{{\bf b}}{2},{\bf p}+\frac{{\bf b}}{2}} &  A_{{\bf p}-\frac{{\bf b}}{2},{\bf p}-\frac{{\bf b}}{2}} \end{array} \right), \label{2bands}
\ee
where $A$ stands for Green functions, self energies, and potentials.
We sum independently the results for every ${\bf b, }$  for which ${\bf b}/2$ is an X-point of the first BZ, and treat  $-{\bf b}$ and  ${\bf b}$ as the same state in order to avoid double counting.
We also consider
$|{\bf p}|$ to be small compared to the lattice momentum, $|{\bf p}| \ll |{\bf b}|$ and expand to quadratic order in ${\bf p}$. The trace in Eq.~(\ref{eq:f3}) is replaced by an integral, $\mbox{Tr}_{\bf p} \to  (1/2) \sum_{{\bf b}} \mbox{Tr}\int \frac{d^3p}{(2\pi)^3}$, where hereinafter trace, Tr$(\cdots)$, is defined only over the $2\times 2$ band space introduced in Eq.~(\ref{2bands}).  

To preform the iteration scheme of Eq.~(\ref{iterations}), 
we use the following definitions:
\begin{eqnarray}
&&\hat{\mathbb U}= \left( \begin{array}{cc} 0 & \mathbb U \\ \mathbb U^* &0 \end{array} \right),\\ 
&& \hat{H}= v_F p_\parallel \left( \begin{array}{cc} 1 & 0 \\ 0 &-1 \end{array} \right) + \left( \begin{array}{cc} 0 & \mathbb U \\ \mathbb U^* &0 \end{array} \right)+ \frac{p_\perp^2}{2m_*}\left( \begin{array}{cc} 1 & 0 \\ 0 &1 \end{array} \right),\nonumber
\end{eqnarray}
where,  $p_\parallel$ and $p_\perp$ are the momenta along and perpendicular to ${\bf b}$ direction, respectively, $v_F$ is the Fermi velocity, and $m^*$ determines the curvature of the parabolic spectrum in the perpendicular direction, see the inset of Fig.~1. As $\delta \hat{n}_2$ does not contain a zero mode contribution, one can use the relaxation time approximation $\mbox{St}_{ee}\{\hat{n}_{eq}+ \delta \hat{n}_2 \}= -\delta \hat{n}_2/\tau_{ee}$. Then taking into account that $\mbox{St}_{U}\{\hat{n}_{eq}+ \delta \hat{n}_1 \}=0$ (proof will be given later) we obtain from Eq.~(\ref{iterations}) the solution for the second iteration:
\begin{subequations}
\label{dn2}
\be
\delta\hat{n}_2=\Upsilon \left( \begin{array}{cc} 2|\mathbb U|^2 & -\mathbb U (\frac{i}{\tau_{ee}}+ 2\epsilon_\parallel ) \\
\mathbb U^* (\frac{i}{\tau_{ee}}- 2\epsilon_\parallel)& -2|\mathbb U|^2  \end{array} \right), 
\ee
where
\be
\Upsilon = \frac{{\bf v}\cdot {\bf b} \tau_{ee}^2\partial_\epsilon n_0(\epsilon)}{1\!+\! 4\left(|\mathbb U|^2+ \epsilon_\parallel^2\right)\tau_{ee}^2}, 
\ee
and $\epsilon_\parallel= v_F p_\parallel$.
\end{subequations}

Substituting this result in Eq.~(\ref{eq:f3}) and calculating the
integral over the momentum, taking into account that $T \gg
1/\tau_{ee}$, we obtain the friction force:
\begin{subequations} 
 \label{eq:ff}
\be {\bf f}^U=-\frac{1}{8\pi} \sum_{{\bf
    b}}({\bf v}\cdot {\bf b}) {\bf b} \frac{ m_*}{v_f}|\mathbb U|^2
F(|\mathbb U |\tau_{ee}), \ee 
where 
\be F(|\mathbb U
|\tau_{ee})= \int_0^\infty d \epsilon \frac{\partial_\epsilon
  n_0(\epsilon)}{\sqrt{1+ 4|\mathbb U|^2 \tau_{ee}^2}} .  \ee
\end{subequations}  
Hereinafter $\tau_{ee}$ should be understood as function of $T$ and
$\epsilon-\mu$.  In this formula the chemical potential is measured
with respect to the zero energy level shown in Fig~1.  This result together
with Eqs.~(\ref{OR2}) and (\ref{eq:rho}) yields the resistivity: 
\be
\rho= \left(\frac{2\pi \hbar}{e^2}a\right)\times \left(\frac{ 2\pi m_* |\mathbb
    U|}{\hbar^2|b|^2} \right)^2\frac{2\pi}{(n a^3)^2}F(|\mathbb U|
\tau_{ee}), \label{main} 
\ee 
where $a=2\pi/|b|$ is the lattice
constant. The first factor on the r.h.s of 
\req{main} has the meaning of the ``maximal metallic resistivity''.
The next factor is the main result of the paper, it shows that at low
temperature, when electron-electron scattering rate is small
$1/\tau_{ee} \ll |\mathbb U|$, $F(|\mathbb U| \tau_{ee})\propto
1/(|\mathbb U| \tau_{ee})$. Namely, in this limit, Eq.~(\ref{main}) recovers the known result for the resistivity due to umklapp scattering, which is proportional to inelastic scattering rate. In the opposite limit, when
electron-electron scattering rate is large, $1/\tau_{ee} \gg |\mathbb
U|$, the function $F(|\mathbb U| \tau_{ee})=1$ for $|\mu|\ll T$,
i.e. at high temperature, the resistivity saturates to a finite value.
[This result is specific for three dimensional systems. In two
dimensions, similar considerations give that the high temperature
resistivity is proportional to $1/\sqrt{T}$ for $|\mu| \ll T$.] Notice that the temperature 
dependence of the resistivity  neither follows the  Matthiessen's rule nor the parallel conductance model
(the latter may be used as a crude fit to \req{main} in a limited temperature range). 

Finally, it is instructive to discuss the violation of the Weidemann-Franz law (WFl) in this regime. Using the standard 
estimate for the thermal conductivity within the hydrodynamic description, we find the Lorenz number
\be
{\cal L}\equiv \frac{\kappa\rho}{T}\simeq \left(\frac{\pi^2}{3e^2}\right)\times 
 \left(\frac{m_* |\mathbb
    U|^2\tau_{ee}}{\hbar^3|b|^2} \right)F(|\mathbb U|\tau_{ee}).
\label{Lorenz}
\ee
(We do not write the numerical factor in the last formula.) 
The first factor on the r.h.s of \req{Lorenz} is the universal number for metals at low temperatures where all the kinetics are determined by the impurity scattering. The remaining factor describes the violation of WFl 
in our theory. For $1/\tau_{ee} \ll |\mathbb U|$ the violation occurs only by the model specific factor, 
whereas in the regime of the resistivity saturation $1/\tau_{ee} \gg |\mathbb U|$, the Lorenz factor acquires 
the  temperature dependence of the electron-electron relaxation time.

To justify the above calculation,  one should show that our solution for 
$\delta \hat{n}_2$ gives rise only to a small renormalization of  the velocity field ${\bf v}$.  To calculate its renormalized value, ${\bf v}_*$, we replace ${\bf v}$ by ${\bf v}_*$ in the expressions to $\delta\hat{n}_1$ and  $\delta\hat{n}_2$ and  substitute them in the constraint (\ref{eq:nv}). The result is
\be
{\bf v}= {\bf v}_* \left[1-\sum_{\bf b}\frac{ {\bf b}\otimes  {\bf b}}{4\pi n} \frac{m_* |\mathbb U|^2}{v_f} \int_0^\infty d\epsilon \frac{ \tau_{ee} \partial_\epsilon n_0(\epsilon)}{\sqrt{1+ 4|\mathbb U|^2 \tau_{ee}^2}}\right]
\ee
The second term in the brackets is at least as small as  $\frac{ m_* |\mathbb U|}{\hbar^2|b|^2}$, and therefore can be neglected.

Finally, to show that St$_{U}\{\hat{n}_{eq}+ \delta \hat{n}_1\}=0$, let us consider the general term: St$_{U}\{\hat{n}_{eq}+ \delta \hat{n}\}$. The diagrammatic expansion of the correction to the potentials associated with the correction to the distribution function $\delta \hat{n}$ is shown in Fig~2b, with the result:
 \begin{eqnarray}
&& \left[ \begin{array}{c} \delta\hat{U}^R(\epsilon)-\delta \hat{U}^A(\epsilon) \label{dUs}\\ \delta \hat{U}^K(\epsilon) \end{array} \right]_{{\bf p},{\bf p}'} \\&&= \frac{1}{Z^2\nu} \int \frac{ d^dq}{(2\pi)^d} \int \frac{ d\epsilon_1}{2\pi} \gamma\left(\frac{\epsilon-\epsilon_1}{v_f|{\bf q}|}\right) \left[ \hat{G}^R \delta \hat{n}- \delta \hat{n} \hat{G}^A\right]_{{\bf p},{\bf p}'} \left[ \begin{array}{c} 1\\ \coth\left( \frac{\epsilon-\epsilon_1}{2 T} \right) \end{array} \right].\nonumber
\end{eqnarray}
Here $\nu$ is the density of states, $\gamma(z)$ is an odd function, $z \gamma (z) \geq 0$, which depends on details of the system and cannot be found from general consideration, and ${\bf p} \neq {\bf p}'$, because the ${\bf p} = {\bf p}'$ terms belong to the self energies $\Sigma^{R,A,K}$ and affect only the zero mode which is irrelevant here. Substituting $\delta \hat{n}=\delta \hat{n}_1$ from Eq.~(\ref{first}) in the above equation shows that $\delta \hat{U}_1^{R,A,K}=0$. Noticing also that $\{\hat{U}^A-\hat{U}^R;\delta\hat{n}_1\}=0$ we obtain: $\mbox{St}_{U}\{\hat{n}_{eq}+ \delta \hat{n}_1 \}=0$.

\section{Subleading corrections to the resistivity}
From the above calculation it follows that the leading contribution to $\mbox{St}_{U}\{\hat{n}_{eq}+ \delta \hat{n} \}$ comes  from $\delta \hat{n}=\delta \hat{n }_2$. Substituting Eq.~(\ref{dn2}) for $\delta \hat{n}_2$ in Eq.~({\ref{dUs}), one obtains:
 \begin{eqnarray}
&& \mbox{St}_U\!\{\hat{n}_{eq}+ \delta \hat{n}_2 \} \!=\! \frac{ {\bf b} \cdot{ \bf v}}{8\pi} \left( \begin{array}{cc} 0  & i \mathbb U \\ - i \mathbb U^* &0 \end{array} \right) \frac{m_*}{\nu v_f} \label{Stu2} \\
&&\times \int_0^\infty\frac{ d\epsilon_1 \partial_{\epsilon_{1}} n_0(\epsilon_1)}{\sqrt{1+ 4 |\mathbb U|^2 \tau_{ee}^2}}\int \frac{d\varphi}{2\pi}
  \gamma\left( \frac{ \epsilon-\epsilon_1}{ v_f| {\bf p}_\perp -{\bf l}_\varphi\sqrt{2m_* \epsilon_1}|}\right) \left[ \coth \left( \frac{\epsilon-\epsilon_1}{2T}\right)
 \! -\!n_{0}(\epsilon) \right],\nonumber
 \end{eqnarray}
where ${\bf l}_\varphi$ is a unit vector in direction of the polar angle $\varphi$ (with ${\bf l}_\varphi \cdot {\bf b}=0$).
Notice that we have neglected the $\{\hat{U}^A-\hat{U}^R;\delta\hat{n}_2\}$ term in $\mbox{St}_u\{\hat{n}_{eq}+ \delta \hat{n}_2 \}$, see Eq.~(\ref{stu}), as it produces only small non-resonant terms.  From Eqs. (\ref{Stu2}) and (\ref{iterations}), one obtains $\delta n_3$ which is substituted in Eq.~(\ref{eq:f3}) to  give the correction to  the friction force:
 \begin{eqnarray}
 \delta {\bf f}^{U}&=&
- \sum_{{\bf b}} ({\bf v}\cdot {\bf b}){\bf b}  
\frac{|\mathbb U|^2m_*}{8\pi v_f} \frac{ m_*}{2\pi \nu v_f}
 \int_0^\infty
\frac{ d\epsilon_{1} d \epsilon_2}{2T\sinh\left(\frac{\epsilon_1-\epsilon_2}{2T} \right)} \label{deltaf} \\
&\times& \int \frac{d\varphi}{2\pi} 
 \gamma\left( \frac{\epsilon_1-\epsilon_2}
{\sqrt{2m_*} v_f|\sqrt{\epsilon_1}{\bf l}_0-\sqrt{\epsilon_2} {\bf l}_\varphi|}\right)
\prod_{j=1,2} 
 \frac{1}{\cosh\left(\frac{\epsilon_j-\mu}{2T}\right)\sqrt{1+ 4 |\mathbb U|^2 \tau^2_{ee}(\epsilon_j)}}. \nonumber
 \end{eqnarray} 
Evaluation of this integral requires knowledge of $\gamma(z)$. The simplest case is that of Fermi liquid theory where $\gamma(z) =\alpha z$ with $\alpha$  a constant of order unity which depends on the interaction strength. In this case we obtain that in the high temperature limit, where  $1/\tau_{ee} \gg |\mathbb U|$, the correction to the resistivity is:
\be 
\delta \rho \simeq\ c \rho_{sat} \left(\frac{T m_*}{\hbar^2 b^2} \right)^{3/2}, \label{subleading} \ee 
where $c$ is a positive constant of order unity, and $\rho_{sat}$ is the saturation value of the resistivity obtained from Eq.~(\ref{main}). At low temperatures, $1/\tau_{ee} \ll |\mathbb U|$,  the correction to the resistivity is small compared to the leading order result Eq.~(\ref{main}) by a factor of order $1/(|\mathbb U|\tau_{ee})$. 

Equation (\ref{subleading}) describes the leading residual temperature
dependence. It is controlled, however, by a much
larger temperature scale than that of main term in
Eq. (\ref{main}). Other sources of the temperature dependence of
the resistivity are the temperature dependence of the potential, $\mathbb U(T),$ due to
screening, and corrections coming from more accurate evaluation of $\mbox{St}_{ee}\{\hat{n} \}$ beyond
the relaxation time approximation. However, these effects
can be shown to be of higher order: $\left(\frac{T m_*}{\hbar^2 b^2} \right)^2$.

\subsection{Semiclassical interpretation}
It is instructive to give a semiclassical
interpretation to the results expressed by Eqs. (\ref{main})
and (\ref{subleading}). The main contribution to the force \rref{def:f} can be
viewed as the oscillatory density perturbation resulting from interference of two electron trajectories
as shown in Fig. 3a (trajectory $1$ is the direct path to the ``observation point'' denoted as $\nabla U$ and trajectory $2$ scattered by the
periodic potential $U(r)$. 
If one thinks that this
interference is limited by any inelastic processes, a simple calculation leads
to Eq.~(\ref{main}) for $U\tau_{ee}\ll 1$. 

There is, however, a finite
set of inelastic amplitudes which preserve the phase coherence and, thus, still contribute to the force. 
The example of such amplitudes is shown in Fig.~3b. In this case the same electron-hole pair is emitted the on both paths $1$ and $2$. As the final states 
are the same, those multi-particles amplitudes also contribute to the interference.  
Evaluating the phase volume
for such processes, while accounting for the enhancement due 
to small angle scattering, gives the correction (\ref{subleading}).

\begin{figure}[t]
\includegraphics[width=0.8\columnwidth]{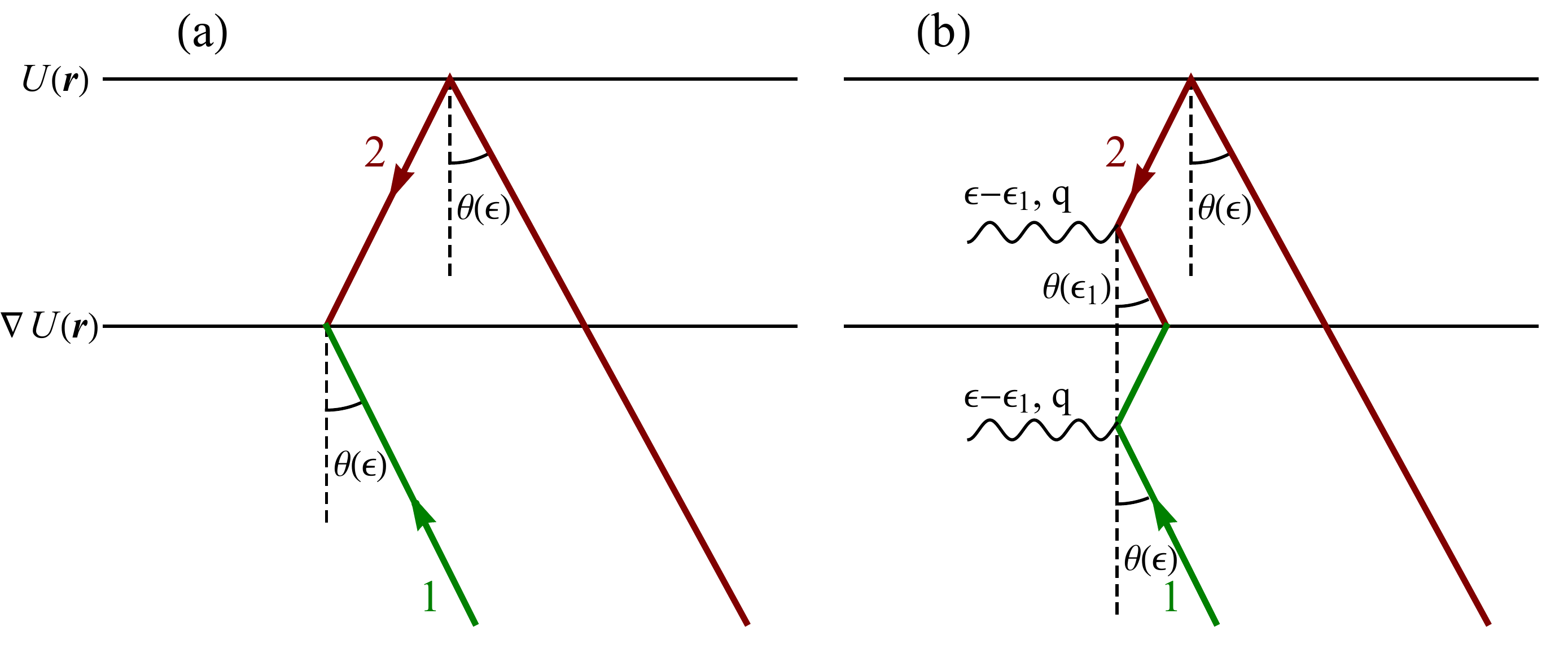}
\caption{Sketch of the interference processes giving rise to the
force: (a) Without inelastic scattering, (b) with small angle
inelastic scattering. (wavy line represents electron-hole excitation).
The angles $\theta(\epsilon)$ are fixed by the resonance condition.}
\end{figure}

\section{Extensions}
\subsection{Other special points in the Brillouin zone:} The results
(\ref{main}) and (\ref{subleading}) were derived for case of two bands, however
their applicability is more general. Indeed, let us choose $\mu >0;~\mu \gg T$ in Eqs. (\ref{eq:ff}) and (\ref{deltaf}). In this case the main contribution is determined by the "hot line" ${\bf p}_\perp^2/(2m^*) \sim \mu \pm T$ , rather than a "hot spot" for $|\mu| \ll T$, see Figs.~1c and 1d.
 This line is the intersection of two closed Fermi surfaces on the  boundary of the BZ. In this case $F(x \to 0)=2$ while in Eq. (\ref{subleading})  one has to replace $T^{3/2}$ by $T^2 \mu^{-1/2}$. 
The insensitivity of the saturation to the shape of the  "hot line" means that the theory  describes  well the intersection of  Fermi surface at  M-points as well as other special symmetry points, see Fig.~1a. (The point itself is of no importance in this case, as all the effect originates from the "hot line".)     

\subsection{The effects of thermal phonons and disorder}
For a model where disorder is realized by shift of the atoms
from their original lattice positions, $\delta {\bf R}_j$, which results in randomization of the phase of the periodic potential, $e^{i\bf{q}\cdot \bf{r}} \to e^{i{\bf q}\cdot{\bf r}+i\delta\phi_{rand}({\bf r})}$,
the high temperature limit of the resistivity (\ref{main}), remains
unaffected provided $v_F\tau_{ee}$ is shorter than the correlation
length of the disorder potential. As shifts, $\delta {\bf R}_j$, in the positions
of the atoms can be produced  either by disorder
or by phonons, this condition  seems to be
consistent with Lindemann's criterion for melting. (Assumptions
in Refs. \cite{Millis} and \cite{Kivelson} about possible destruction
of the band structure by thermal phonons seem to
explicitly contradict the Lindemann criterion\cite{footnote2}.) At lower
temperature (with $v_F \tau_{ee}$ longer than the elastic mean
free path) disorder leads to finite resistivity which does
not depend anymore on umklapp scattering.

\section{Conclusion} We have constructed a controllable theory of the resistivity saturation based on  electron-electron interaction rather than the conventional attempts to describe the effect as resulting from phonon destruction of the energy band structure. It is leading order in $|\mathbb U|/T$, correct to all orders in $|\mathbb U| \tau_{ee}$. From this theory it follows that in 3D systems the resistance saturates to a constant value with subleading correction proportional either to $T^{3/2}$ or $T^2$ depending whether the Fermi surface touches the boundary of the BZ at a "hot spot" ($|\mu| \ll T$)  or a "hot line" ($\mu \gg T$), respectively. Similar calculation for 2D materials shows that the high temperature resistance decreases as $T^{-1/2}$ with subleading corrections proportional to $T^{1/2}$ when $|\mu| \ll T$,
and saturated resistivity with  $T^2$  correction for  $\mu \gg T$.

The main small parameter of the theory is $\delta
=\frac{|\mathbb U| m_*}{2\pi\hbar^2 b^2}$, which is of the order of the ratio of band splitting to the band width. 
In A15 compounds, a typical value of the saturation resistance 
is $\rho_{sat} \sim 150 \mu \Omega \mbox{cm}$, while the lattice constant is $a\sim 5$ \AA \cite{A15Lattice}.  Then from Eq.~(\ref{main}), for $na^3=1$, it follows that $\delta\sim 0.14$, which does not look unrealistic (an additional enhancement of the resistance
is provided by the larger number of anti-crossing bands. We will not dwell on this issue here as it requires more reliable knowledge of the details of the band structure).

\section*{Acknowledgment}

We acknowledge  discussions of the results with A.V. Andreev, Yu.M. Galperin, L.I. Glazman, A.J. Millis, and Z. Ovadyahu.
This research was supported by the Israel Science Foundation (ISF) Grant No. 302/14 (O.A.), and by the Simons Foundation (I.A.).

\appendix
\section{Hydrodynamics of non-Galilean invariant systems}
 
The non-dissipative part of the hydrodynamic description is readily illustrated under the assumption that the kinetic equation (KE) is applicable, even though the structure of hydrodynamics does not require its validity. In the absence of umklapp scattering, electron distribution function of the form
\be
f_{\mu,T, {\bf v}} = f_F( \varepsilon({\bf k})- {\bf v} \cdot {\bf k})~ ;~~~~ f_F(\varepsilon)= \frac{1}{1+ e^{\frac{\varepsilon-\mu}{T}}} \label{A1} 
\ee  
is a solution of the KE.  Here $\varepsilon({\bf k})$ is the  energy spectrum of the electrons, $T$ is the temperature, $\mu$ is the chemical potential, and ${\bf v}$ is some arbitrary velocity. Five independent parameters of this distribution function express the conservation of energy ($T$), charge ($\mu$) and the three components of the momentum ${\bf v}$.   The general form of the distribution function (\ref{A1}) is fixed by the strong inelastic electron-electron collisions. It does not imply Galilean invariance.  
 
The current density is given by
\begin{eqnarray}
 j_\alpha &=& e \int \frac{d^3 k}{(2\pi)^3} \frac{\partial \varepsilon({\bf k})}{\partial k_\alpha}f_{\mu,T,{\bf v}} 
\\ &=&  e v_\alpha \int \frac{d^3 k}{(2\pi)^3}f_{\mu,T,{\bf v}}+e \int \frac{d^3 k}{(2\pi)^3} \frac{\partial \varepsilon({\bf k})-{\bf v} \cdot {\bf k}}{\partial k_\alpha}f_{\mu,T,{\bf v}}  \nonumber
\\&=& n ev_\alpha+ \mbox{(boundary term)$_\alpha$} \nonumber
\end{eqnarray}
Here
\be 
n=  \int \frac{d^3 k}{(2\pi)^3}f_{\mu,T,{\bf v}}
\ee
is the particle density, while 
\be
 \mbox{(boundary term)$_\alpha$}= e T \int \frac{d^3 k}{(2\pi)^3} \frac{ \partial}{\partial k_\alpha} \ln \left( 1-f_{\mu,T,{\bf v}} \right) \label{A4}
\ee
Thus if the boundary term can be neglected 
\be 
j_\alpha = en v_\alpha ~~~~~~\mbox{(electron bands)} \label{A5}
\ee
independent of the spectrum of the system, $\varepsilon ({\bf k})$.

The derivation of Eq.~(\ref{A5}) requires the boundary term (\ref{A4}) to vanish. This is always the case when the integration over ${\bf k}$ is unrestricted, and $\varepsilon ({\bf k})$ is an arbitrary function which satisfies   $\varepsilon (|{\bf k}| \to \infty) \to \infty$.  However, in crystals, $\varepsilon ({\bf k})$ is limited within the BZ, therefore the requirement is that
\be
f_{\mu,T,{\bf v}}({\bf k} \in \mbox{BZ boundary}) \to 0. ~~~\mbox{(electron bands)} \label{A6}
\ee

This condition is satisfied when (a) the band is partially field; (b) the width of the band of the Hamiltonian $H_h$  is larger than $T$.  The same condition (\ref{A6}) removes the apparent inconsistency: $
f_{\mu,T,{\bf v}}({\bf k}+{\bf b})\neq f_{\mu,T,{\bf v}}({\bf k})$. (The jump on the boundary of the BZ is exponentially small).

The second possibility for removing the boundary term (\ref{A4}) is 
\be
f_{\mu,T,{\bf v}}({\bf k} \in \mbox{BZ boundary}) \to 1. ~~~\mbox{(hole bands)} \label{A7}
\ee 
 This can be achieved by the electron-hole transformation $f_{\mu,T,{\bf v}} \to 1-f^{(h)}_{\mu, T,{\bf v}}$, ${\bf k} \to - {\bf k},~~ \varepsilon({\bf k}) \to  - \varepsilon({\bf k}),~~ \mu \to - \mu$, and we obtain: 
\be
j_\alpha = - e n v_\alpha;~~~~~~~   n=\int \frac{d^3 k}{(2\pi)^3}f^{(h)}_{\mu,T,{\bf v}}~~~~ \mbox{(hole bands)} 
\ee
 
Finally, we emphasize that the BZ of the Hamiltonian $H_h$ is twice as large than that of the total Hamiltonian, and assumptions (\ref{A6}) and ({\ref{A7})  are consistent with further calculations. 
 
 Assuming the absence of boundary terms  we find for the momentum density \begin{eqnarray}
p_\alpha &=& \int \frac{ d^3 k}{(2\pi)^3} k_\alpha f_{\mu,T,{\bf v}} = \frac{\partial}{\partial v_\alpha} T  \int \frac{ d^3 k}{(2\pi)^3} \ln \left( \frac{1}{1-  f_{\mu,T,{\bf v}}}\right) \nonumber\\
&=&  \frac{\partial}{\partial v_\alpha} P(\mu, T, {\bf v})
\end{eqnarray}
where the last equality follows from the usual grand canonical definition of the pressure ($ P(\mu, T, {\bf v})=  P(\mu, T, -{\bf v})$ by time reversal symmetry).  It holds for both for interacting and noninteracting systems. Similarly, the thermodynamic relations:
\be
n= \frac{\partial}{\partial \mu} P(\mu, T, {\bf v})~~~~ s= \frac{\partial}{\partial T} P(\mu, T, {\bf v})
\ee
also hold.
 
For Galilean invariant systems: 
\be
 P(\mu, T, {\bf v})= P(\mu+ \frac{ m {\bf v}^2}{2}, T) \label{A11}
\ee
so that
$p_\alpha = m n v_\alpha$ where $m$ is the electron mass (not renormalized by electron-electron interactions).  For non-Galilean invariant systems (\ref{A11}) no longer holds but in the limit ${\bf v} \to 0$,
\be
p_\alpha= n m^*_{\alpha,\beta}(n,T) v_\beta,
\ee
where the symmetric effective mass tensor, $m^*_{\alpha,\beta}(n,T)$, determined by the density, temperature and interaction is not related to the mass extracted from compressibility or specific heat. 

Finally one can derive the expression for the energy current:
\begin{eqnarray}
j^\alpha_\epsilon &=& \int \frac{ d^3 k}{(2\pi)^3} \varepsilon({\bf k}) \frac{\partial\varepsilon({\bf k})}{\partial k_\alpha} f_{\mu,T,{\bf v}}
\\
& = & \int \frac{ d^3 k}{(2\pi)^3}\left[ \varepsilon({\bf k})- \mu - {\bf v} \cdot {\bf k} \right] \left[\frac{\partial\varepsilon({\bf k})}{\partial k_\alpha} - v_\alpha \right]f_{\mu,T,{\bf v}}\nonumber\\
&+& v_\alpha\int \frac{ d^3 k}{(2\pi)^3}\left[ \varepsilon({\bf k})- \mu - {\bf v} \cdot {\bf k} \right]f_{\mu,T,{\bf v}} \nonumber\\
&+&  \int \frac{ d^3 k}{(2\pi)^3}\left[ \mu + {\bf v} \cdot {\bf k} \right] \left[\frac{\partial\varepsilon({\bf k})}{\partial k_\alpha} - v_\alpha \right]f_{\mu,T,{\bf v}}\nonumber\\
&+&  v_\alpha \int \frac{ d^3 k}{(2\pi)^3}\left[ \mu + {\bf v} \cdot {\bf k} \right] f_{\mu,T,{\bf v}}\nonumber
\end{eqnarray}
The first term is a total derivative which vanishes under the assumption discussed above, and for the remaining terms one finds:
\be
j_\epsilon^\alpha= v_\alpha T \frac{ \partial P}{\partial T}+ v_\alpha P+ v_\alpha (v_\beta p_\beta)+ \mu n v_\alpha.  \label{A14}
 \ee
 The first term in this expression is nothing but the thermal current $T q_\alpha$ where $q_\alpha = s v_\alpha$. Once again, Eq.~(\ref{A14}) does not require Galilean invariance. It relies on the strong inelastic electron-electron scattering  which stabilizes the distribution function (\ref{A1}).


\begin{thebibliography}{99}
\bibitem{footnote2} According to the empirical Lindemann criterion, the displacement fluctuations are much smaller than the lattice constant even at the melting temperature,  i.e.  $\langle u^2 \rangle < 0.01 a^2$  (this is simply an indication of the fact that liquid-solid phase transition is a strong first order transition). Phonons play the role of static disorder, and one can evaluate the electron transport properties limited by such atoms displacements both in the strong (tight binding) and in the weak coupling limits. It shows that the metal insulator transition point is unreachable in both cases, namely all the metallic results for three dimensional systems  still hold. We believe that the results of Refs. 21 \& 22 are artifacts of (a) using unphysical limit of infinitely large electron-phonon interaction constant, i.e. $\langle u^2 \rangle > a^2$,  and (b) disregarding the localization physics even in this unphysical regime.
\end{thebibliography}
\end{document}